\documentclass[aps,showpacs,prd,onecolumn]{revtex4}%
\usepackage{graphicx}
\usepackage{amssymb}
\usepackage{amsmath}
\usepackage{color}
\usepackage{float}
\usepackage{tabularx}
\usepackage{ulem}
\usepackage{accents}
\usepackage{graphicx}
\usepackage{amsfonts}

\usepackage{makecell}

\usepackage[colorlinks=true,pdfstartview=FitV,linkcolor=blue,citecolor=blue,urlcolor=blue,breaklinks=true]%
{hyperref}%
\providecommand{\U}[1]{\protect\rule{.1in}{.1in}}
\begin{document}
\title{Constraining dimension-six nonminimal Lorentz-violating electron-nucleon
interactions with EDM physics}
\author{Jonas B. Araujo$^{a,b}$}
\email{jonas.araujo88@gmail.com}
\author{A. H. Blin$^{c}$}
\email{alex@uc.pt}
\author{Marcos Sampaio$^{b,d}$}
\email{marcos.sampaio@ufabc.edu.br}
\author{Manoel M. Ferreira Jr$^{a}$}
\email{manojr.ufma@gmail.com}
\affiliation{$^{a}$ Departamento de F\'{\i}sica, Universidade Federal do Maranh\~{a}o,
Campus Universit\'{a}rio do Bacanga, S\~{a}o Lu\'{\i}s - MA, 65080-805 - Brazil}
\affiliation{$^{b}$ Centre for Particle Theory, University of Durham, Durham, DH1 3LE, United Kingdom}
\affiliation{$^{c}$ CFisUC, Department of Physics, University of Coimbra, P-3004-516 Coimbra, Portugal}
\affiliation{$^{d}$ CCNH, Universidade Federal do ABC, 09210-580 , Santo Andr\'e - SP, Brazil}

\begin{abstract}
	The electric dipole moment (EDM) of an atom could arise also from
	$P$-odd and $T$-odd electron-nucleon couplings. In this work we investigate a
	general class of dimension-$6$ electron-nucleon ($e$-$N$) nonminimal interactions
	mediated by Lorentz-violating (LV) tensors of rank ranging from $1$ to $4$. The possible couplings are listed as well as their behavior under
	$C$, $P$ and $T$, allowing us to select the couplings compatible with EDM physics. The
	unsuppressed contributions of these couplings to the atom's
	Hamiltonian can be read as EDM-equivalent. The LV coefficients' magnitudes are limited using EDM experimental data to the level of $3.2\times
	10^{-13}\text{(GeV)}^{-2}$ or $1.6\times10^{-15}\text{(GeV)}^{-2}$.

\end{abstract}

\pacs{11.30.Cp, 11.30.Er, 13.40.Em}
\maketitle

\section{Introduction}

Electric dipole moments (EDMs) are excellent probes for violation of discrete
symmetries \cite{EDM1,EDM2,EDM3,LeptonEDM} and of physics beyond the standard
model (SM) \cite{Yamanaka,Pospelov2005}. EDM terms violate both parity ($P)$
and time reversal ($T)$ symmetries, while preserve charge conjugation ($C$),
if the $CPT$ theorem holds. In the SM structure, $P$ and $CP$ violations have
been proposed and detected since the $1950$s, but violations of $T$ only
recently have been measured by the $BABAR$ Collaboration \cite{Tviolation}.
Furthermore, the role of $CP$ violation is crucial in explaining the baryon
asymmetry in the universe, as pointed out by Sakharov \cite{Sakharov}. However, as a standard EDM is $C$-even, it cannot play a role by itself on the baryon asymmetry in the universe.
Concerning the EDM experiments, the growing experimental precision has lead to
stringent upper bounds on several $CP$-violating theories \cite{Measure1,
Measure2,Baron,EDMnature2018}. If an atomic EDM were to be detected, it could
arise from intrinsic properties of the electrons and/or nucleus, or from
$P$-odd and $T$-odd electron-nucleon ($e$-$N$) couplings.

In the SM, the electronic EDM is generated via radiative corrections at four
loop order \cite{SMEDM1,SMEDM2}, with magnitude $d_{e-\text{SM}}$
$\simeq10^{-38}\,e\cdot{\text{cm}}$. In addition, the current experimental
measurement has ruled out an electronic EDM up to $d_{e-{\text{exp}}}$
$=1.1\times10^{-31}\,e\cdot{\text{cm}}$ \cite{EDMnature2018}, which is still 7
orders of magnitude above the SM prediction. Concerning the EDM searches, it
is important to remember that the carrier of EDM may be the electron or the
atomic nucleus. In fact, the atomic nucleus, due to $P$-odd and $T$-odd
nuclear interactions \cite{Flambaum1}, can yield a nuclear EDM, which, for a
pointlike nucleus, would be screened by the very interesting Schiff's theorem
\cite{Schiff}. For a finite-sized nucleus, however, the screening is not
total, and the Schiff moment appears as a residual effect \cite{Flambaum1}.
Furthermore, if relativity is considered, not only the screening is
ineffective, but the whole atom's or molecule's EDM may exceed the electron's
by a few orders of magnitude \cite{Sandars0,Sandars1,Sandars2,Sternheimer2}.
Recent experiments to measure nuclear EDM have been undertaken and an upper
bound of $d_{^{199}\text{Hg}}<7.4\times10^{-30}\,e\cdot{\text{cm}}$ was set
for the $^{199}\text{Hg}$ nucleus \cite{Heckel}. The surprisingly minute
nuclear EDM, which is related to the QCD $\theta$-term, poses the strong $CP$
problem, whose solution could involve the yet undetected axions \cite{Axion}.

 There are three ways to generate EDM for an atom. Two of them rely on EDM sources in the atom's constituents and appropriate means of circumventing the Schiff's theorem: (i) relativistic enhancement of the electron EDM; (ii) partial screening of the nuclear EDM due to the finite-sized nucleus. The third mechanism is based on $P$-odd and $T$-odd electron-nucleon interactions. Thus, even in an atom deprived of particle EDM sources  -- (i) and (ii), an effective atomic EDM
may be engendered via dimension-$6$ interactions between the nucleus and the
electronic cloud, being represented by $P$-odd and $T$-odd electron-nucleon
($e$-$N$) couplings, such as%
\begin{equation}
\mathcal{L}_{CP}=i\frac{G_{F}}{\sqrt{2}}\sum\limits_{j}\left[  C_{S}\bar
{N}_{j}N_{j}\bar{\psi}\gamma^{5}\psi+C_{P}\bar{N}_{j}\gamma^{5}N_{j}\bar
{\psi}\psi+C_{T}\bar{N}_{j}\sigma^{\mu\nu}N_{j}\bar{\psi}\sigma^{\mu\nu
}\gamma^{5}\psi\right]  \label{UsualCase}.
\end{equation}

Among these, the dominant contribution arises from the scalar-pseudoscalar
coupling, $C_{S}\bar{N}_{j}N_{j}\bar{\psi}\gamma^{5}\psi$, where $G_{F}$
denotes Fermi's constant \cite{SMEDM2,LeptonEDM,AtMolPhys,Barr}. By supposing
that the $e$-$N$ couplings are the sole source of EDM for the atom, the
dimensionless coefficient $C_{S}$ can be constrained to the level
$C_{S}<7.3\times10^{-10}$, according to the most precise experiment up to date on the electron's EDM \cite{EDMnature2018}. Moreover, the
scalar-pseudoscalar coupling is similar to the dominant term in ordinary
atomic parity nonconservation (PNC), originated from the coupling of the axial
electronic neutral weak current to the vector nucleonic neutral weak current
via a $Z^{0}$ exchange \cite{LeptonEDM, AtMolPhys, Bouchiat}. The effects on atomic polarization in heavy atoms have also been studied \cite{PNC2,EDM2,AtPol}. $P$-odd and
$T$-odd interactions in atomic systems may yield non-null matrix elements in
heavy atoms with one valence electron \cite{Neuffer}.  Other $e$-$N$
couplings, including the tensor-pseudotensor and pseudoscalar-scalar, are
investigated via atomic calculations for the $^{199}\text{Hg}$ nucleus
\cite{YamanakaHg}.

EDM phenomenology can also arise in a Lorentz-violating (LV) scenario,
addressed within the framework of the Standard Model extension (SME),
developed by  Kostelecky, Colladay and others in Refs. \cite{Colladay}. The SME includes
dimension-$4$ and dimension-$3$ LV terms in all sectors of the SM, comprising
fermions \cite{fermion,CPT,fermion2,fermion3}, photons \cite{KM1,CFJ,photons1}, Yang-Mills developments \cite{YM}, Casimir effect \cite{Casimir}, 
photon-fermion interactions \cite{Vertex,QED} and electroweak (EW) processes
\cite{EW1,EW2,EW3}. The minimal SME can be further extended so as to contain
nonminimal couplings composed of higher-order derivatives in the photon \cite{NMSME1} and in the fermion sector \cite{NMSME2}, as well as higher-dimension operators \cite{Reyes,HD,Ding}.  Nonminimal couplings deprived of higher-order derivatives (except the one contained in the field strength) have been proposed in describing LV interactions between fermions and photons \cite{NModd1,NModd2,PetrovPLB2016} and LV interactions in the electroweak sector \cite{Victor}.
Dimension-5 terms of Myers-Pospelov
	type have also been investigated in the context of black-body radiation \cite{Anacleto}
	and emission of electromagnetic and gravitational waves \cite{Anacleto2}.

Lorentz violation can also work as a source of $CP$ violation and EDM
via radiative corrections \cite{Haghig} or even at tree level via dimension-$5$ nonminimal couplings \cite{Pospelov2008,FredeNM1,Jonas1,Jonas2}. Lorentz-violating (LV)
dimension-$5$ nonminimal couplings have been proposed as nonusual QED
interactions between fermions and photons, yielding EDM Lagrangian pieces as
$\lambda\bar{\psi}(K_{F})_{\mu\nu\alpha\beta}\Gamma^{\mu\nu}F^{\alpha\beta
}\psi$ and $\lambda_{1}\bar{\psi}T_{\mu\alpha}F^{\alpha}{}_{\nu}\Gamma^{\mu\nu
}\psi$, where $(K_{F})_{\mu\nu\alpha\beta}$\ and $T_{\mu\alpha}$ are
$CPT$-even LV tensors, and $\Gamma^{\mu\nu}$ are combinations of Dirac matrices \cite{FredeNM1,Jonas1,Jonas2}. Electron
EDM experimental data has yielded upper bounds as tight as $10^{-25}%
\,(\mbox{eV})^{-1}$ on the magnitude of these couplings. Looking at another route, nuclear EDMs may also be connected with LV physics. Lorentz-violating contributions to the nuclear Schiff moment have been investigated as well \cite{Jonas3}. It is worth mentioning that LV
theories were also analyzed in connection with the magnetic dipole moment (MDM) physics \cite{Stadnik, Gomes, Haghig2}, notwithstanding providing less severe upper bounds.

In this work, we investigate a class of dimension-$6$ and Lorentz-violating
$e$-$N$ couplings, composed by rank-$1$, rank-$2$, rank-$3$, and rank-$4$ background
	tensors, and the possible generation of atomic EDM. These couplings
were first proposed in a recent generalization of gauge theories with LV operators
of arbitrary dimension \cite{LVArbDim}, which contains nonminimal couplings of dimensions ranging from $5$ to $8$. Specifically, we are interested in
the dimension-$6$ fermion-fermion interactions displayed here in Tables \ref{TableI} and \ref{Table4}, for the purpose of generating EDM. This work is organized as follows. In Sec. \ref{section2},
several possibilities of dimension-$6$ couplings are presented and
analyzed concerning their behavior under $C$, $P$, and $T$ operations and
suppression criteria for yielding EDM. Also, redundancies in the couplings are illustrated and commented. In Sec. \ref{section3}, we
examine the Hamiltonians corresponding to unsuppressed EDM couplings, and their
respective energy shifts are estimated and limited using the current
experimental data. The sidereal analysis on the LV terms is also
performed. In Sec. \ref{section4}, the conclusions are drawn.

\section{Nonminimal electron-nucleon Lorentz-violating couplings}\label{section2}

An atomic EDM could be the result of the EDM contained in the electrons or
nucleons, or it could be due to $P$-odd and $T$-odd electron-nucleon
interactions only. Lorentz violation is a natural source of $CP$-breaking and can
work as an environment to generate $P$-odd and $T$-odd $e$-$N$ interactions.
In this sense, we are interested in a class of Lorentz-violating (LV) electron-nucleon
couplings. Considering dimension-$6$ couplings involving $2$ fermions, we are restricted to derivative-free couplings, otherwise these would have dimension higher than $6$, one unit higher for each extra derivative considered. The simplest case involves a rank-$1$ LV
tensor $(k_{XX})_{\mu}$, so that the effective dimension-$6$ Lagrangian piece should
have the form
\begin{equation}
\mathcal{L}_{\text{LV}}=(k_{XX})_{\mu}\left[  \left(  \bar{N}\,\Gamma_{1}\,N\right)
\left(  \bar{\psi}\,\Gamma_{2}\,\psi\right)  \right]  ^{\mu}\ ,
\end{equation}
indicating that the upper index $\mu$ belongs to either $\Gamma_{1}$ or
$\Gamma_{2}$. In addition, the subscript $XX$ in $(k_{XX})_{\mu}$ refers to the nature of the two fermion bilinears as 
follows: scalar ($S$), pseudoscalar ($P$), vector ($V$), axial vector ($A$) and tensor ($T$), which account for the $16$ linearly independent $4\times4$ matrices. Due to the limitation on dimension-$6$ couplings,
the operators $\Gamma_{1,2}$ must be combinations of Dirac matrices. We use the definitions
\begin{equation}
\alpha^{i}=\left(
\begin{array}
[c]{cc}%
0 & \sigma^{i}\\
\sigma^{i} & 0
\end{array}
\right)  ,\ \ \Sigma^{k}=\left(
\begin{array}
[c]{cc}%
\sigma^{k} & 0\\
0 & \sigma^{k}%
\end{array}
\right)  ,\ \ \gamma^{i}=\left(
\begin{array}
[c]{cc}%
0 & \sigma^{i}\\
-\sigma^{i} & 0
\end{array}
\right)  ,\ \ \gamma^{0}=\left(
\begin{array}
[c]{cc}%
1 & 0\\
0 & -1
\end{array}
\right)  ,\ \ \gamma^{5}=\left(
\begin{array}
[c]{cc}%
0 & 1\\
1 & 0
\end{array}
\right)  ,\label{Defsgamma}%
\end{equation}
and $\sigma^{\mu\nu}=i\left[  \gamma^{\mu},\gamma^{\nu}\right]  /2$, with
$\sigma^{0j}=i\alpha^{j}$, $\sigma^{ij}=\epsilon_{ijk}\Sigma^{k}$. As we are
interested in the generation of EDM, we should focus on the
$P$-odd and $T$-odd couplings. In principle, one can have
$\Gamma_{1},\Gamma_{2}=\gamma^{\mu},\gamma^{\mu}\gamma^{5},\gamma^{5},1,$
which provide several possibilities,%
\begin{equation}
\Gamma_{1}=\gamma^{\mu},\Gamma_{2}=1;\text{ }\Gamma_{1}=\gamma^{\mu}%
,\Gamma_{2}=\gamma^{5};\Gamma_{1}=\gamma^{\mu}\gamma^{5},\Gamma_{2}=1;\text{
}\Gamma_{1}=\gamma^{\mu}\gamma^{5},\Gamma_{2}=\gamma^{5};
\end{equation}
and the corresponding combinations interchanging $\Gamma_{1}$ and $\Gamma_{2}%
$, yielding eight couplings that are listed in Table \ref{TableI}. 
\begin{table}[h]%
\begin{tabular}
[c]{|l|c|c|c|c|}\hline
Coupling &  & \ \ \ $P$-odd, $T$-odd piece\ \ \  & NRL & EDM\\\hline
$\left(  k_{SV}\right) _{\mu}(\bar{N}N)(\bar{\psi}\gamma^{\mu}\psi)$ &  &
$\left(  k_{SV}\right) _{i}(\bar{N}N)(\bar{\psi}\gamma^{i}\psi)$ & NS & yes \\\hline
$\left(  k_{VS}\right) _{\mu}(\bar{N}\gamma^{\mu}N)(\bar{\psi}\psi)$ &  &
$\left(  k_{VS}\right) _{i}(\bar{N}\gamma^{i}N)(\bar{\psi}\psi)$ & S & --\\\hline
$\left(  k_{VP}\right)  _{\mu}(\bar{N}\gamma^{\mu}N)(\bar{\psi}i\gamma_{5}\psi)$ &
& $\left(  k_{VP}\right) _{0}(\bar{N}\gamma^{0}N)(\bar{\psi}i\gamma_{5}\psi)$ & NS & yes \\\hline
$\left(  k_{PV}\right)  _{\mu}(\bar{N}i\gamma_{5}N)(\bar{\psi}\gamma^{\mu}\psi)$ &
& $\left(  k_{PV}\right) _{0}(\bar{N}\gamma_{5}N)(\bar{\psi}\gamma^{0}\psi)$ & S & --\\\hline
$\left(  k_{SA}\right)  _{\mu}(\bar{N}N)(\bar{\psi}\gamma^{\mu}\gamma_{5}\psi)$ &  &
none & -- & --\\\hline
$\left(  k_{AS}\right)  _{\mu}(\bar{N}\gamma^{\mu}\gamma_{5}N)(\bar{\psi}\psi)$ &  &
none & -- & --\\\hline
$\left(  k_{PA}\right)  _{\mu}(\bar{N}i\gamma_{5}N)(\bar{\psi}\gamma^{\mu}%
\gamma_{5}\psi)$ &  & none & -- & -- \\\hline
$\left(  k_{AP}\right)  _{\mu}(\bar{N}\gamma^{\mu}\gamma_{5}N)(\bar{\psi}%
i\gamma_{5}\psi)$ &  & none & -- & --\\\hline
\end{tabular} 
\caption{General $CPT$-odd couplings with a rank-$1$ LV tensor and Dirac bilinears. NRL stands for the nonrelativistic limit for the nucleons ($N$). In this limit, the coupling component can be suppressed, ``S'', or not suppressed, ``NS''.} \label{TableI}
\end{table}

$CPT$-odd dimension-$6$ couplings, containing the tensor
operator $\sigma^{\mu\nu}$, can also be proposed. The initial options would be%
\begin{equation}
\Gamma_{1}=\gamma^{\mu},\Gamma_{2}=\sigma^{\mu\nu};\text{ }\Gamma_{1}%
=\gamma^{\mu}\gamma^{5},\Gamma_{2}=\sigma^{\mu\nu};
\end{equation}
to which one adds the corresponding combinations letting $\Gamma
_{1}\leftrightarrow\Gamma_{2},$ engendering four possibilities,
displayed in Table \ref{Table2}. These couplings, however, are included in a rank-$3$ generalization. Observe, for instance: 
\begin{equation}
(\kappa_{VT})_{\nu }(\bar{N}\gamma _{\mu }N)(\bar{\psi}\sigma ^{\mu \nu
}\psi)=(\kappa_{VT})_{\nu }g_{\beta \mu }(\bar{N}\gamma ^{\mu }N)(\bar{\psi}\sigma
^{\beta \nu }\psi)=(k_{VT})_{\nu \beta \mu }(\bar{N}\gamma ^{\mu }N)(\bar{\psi}%
\sigma ^{\beta \nu }\psi)\ , \label{Redundancy1}
\end{equation}%
where $(k_{VT})_{\nu \beta \mu }=(k_{VT})_{\nu }g_{\beta \mu }$ is a
particular parametrization, with $g_{\beta\mu}$ being the Minkowski metric tensor. Consequently, all the rank-$1$ couplings with one contracted index
between the bilinears can be reproduced by the rank-$3$ structures of Table \ref{Table4}. In other words, all the rank-$1$ couplings of Table \ref{Table2} are included as particular cases of the rank-$3$ couplings to be shown in Table \ref{Table4}, which also includes the couplings of rank-$2$ and rank-$4$.

\begin{table}[h]%
\begin{tabular} 
[c]{|l|c|c|c|c|}\hline
Coupling &  & $P$-odd, $T$-odd piece & NRL & EDM\\\hline
$\left(k_{VT}\right)_{\nu} (\bar{N}\gamma_{\mu}N)(\bar{\psi}\sigma^{\mu\nu}\psi)$ &  & $$none & -- &
--\\\hline
$\left(k_{AT}\right)_{\nu}(\bar{N}\gamma_{\mu}\gamma_{5}N)(\bar{\psi}\sigma^{\mu\nu}\psi)$ &  &
$\left(k_{AT}\right)_{0}(\bar{N}\gamma_{i}\gamma_{5}N)(\bar{\psi}\sigma^{i0}\psi)$ & NS &
yes \\\hline
$\left(k_{TV}\right)_{\nu}(\bar{N}\sigma^{\mu\nu}N)(\bar{\psi}\gamma_{\mu}\psi)$ &  & none & -- &
--\\\hline
$\left(k_{TA}\right)_{\nu}(\bar{N}\sigma^{\mu\nu}N)(\bar{\psi}\gamma_{\mu}\gamma_{5}\psi)$ &  &
$\left(k_{TA}\right)_{0}(\bar{N}\sigma^{i0}N)(\bar{\psi}\gamma_{i}\gamma_{5}\psi)$ & S & --\\\hline
\end{tabular}
\caption{Redundant $CPT$-odd couplings with a rank-$1$  LV tensor and matrixes $\gamma^{\mu}$,
$\sigma^{\mu\nu}$ and $\gamma^{5}$}
\label{Table2}
\end{table}

Other combinations involving $\sigma^{\mu\nu}$, such as
\begin{align}
\Gamma_{1} &  =\gamma_{\nu}\sigma^{\mu\nu},\text{ }\Gamma_{2}=1,\\
\text{ }\Gamma_{1} &  =\gamma_{\nu}\sigma^{\mu\nu},\text{ }\Gamma_{2}%
=\gamma^{5},\text{ }\\
\Gamma_{1} &  =\gamma_{\nu}\sigma^{\mu\nu}\gamma^{5},\text{ }\Gamma_{2}=1,
\end{align}
could also be proposed, but dot not bring novelty, due to the
redundancy of the product $\gamma_{\nu}\sigma^{\mu\nu}=3i\gamma^{\mu}.$ In addition, the identity
\[
\sigma _{\mu \nu }\gamma _{5}=\frac{i}{2}\epsilon _{\mu \nu \rho \theta}\sigma ^{\rho \theta },
\]%
where $\epsilon _{\mu \nu \rho \theta}$ (with $\epsilon _{0123}=+1$) is the Levi-Civita symbol, frees us from taking ``pseudotensor" ($\sigma _{\mu \nu }\gamma _{5}$) terms into account, since these can be written in terms of the tensor ones, as follows:
\begin{align}
\left((k_{V-PT})^{\nu }\right)(\bar{N}\gamma ^{\mu }N)(\bar{\psi}i\sigma _{\mu \nu }\gamma
_{5}\psi) =\left( k_{VT}\right) _{\mu \rho \theta }(\bar{N}\gamma ^{\mu }N)(%
\bar{\psi}\sigma ^{\rho \theta }\psi),
\end{align}%
where 
\begin{align}
\left( k_{VT}\right) _{\mu \rho \theta } =\frac{1}{2}\epsilon _{\nu \mu\rho \theta }(k_{V-PT})^{\nu } . 
\end{align}

Thus, these couplings are equivalent to the rank-$3$ ones displayed in Table V.

Concerning the couplings in Tables \ref{TableI} and \ref{Table4}, these are considered suppressed when the nucleon's
bilinear mixes large and small spinor components, becoming negligible in the
nonrelativistic limit. We obviously are interested only in the unsuppressed terms, i.e., the bilinears composed of only large
components. The behavior of the couplings under discrete symmetries depends on
the way the Dirac bilinears transform under these symmetry operations, as shown in
Table \ref{Table3}.
\begin{table}[h]
$%
\begin{tabular}
[c]{|l|c|c|c|c|c|c|c|c|}\hline
& $\bar{\psi}\psi$ & $\bar{\psi}i\gamma_{5}\psi$ & $\bar{\psi}\gamma^{0}\psi$
& $\bar{\psi}\gamma^{i}\psi$ & $\bar{\psi}\gamma^{0}\gamma_{5}\psi$ &
$\bar{\psi}\gamma^{i}\gamma_{5}\psi$ & $\bar{\psi}\sigma^{0i}\psi$ &
$\bar{\psi}\sigma^{ij}\psi$\\\hline
$P$ & $+$ & $-$ & $+$ & $-$ & $-$ & $+$ & $-$ & $+$ \\\hline
$T$ & $+$ & $-$ & $+$ & $-$ & $+$ & $-$ & $+$ & $-$ \\\hline
$C$ & $+$ & $+$ & $-$ & $-$ & $+$ & $+$ & $-$ & $-$ \\\hline
\end{tabular}
$\caption{Behavior of Dirac bilinears under discrete symmetry operators.} \label{Table3}
\end{table}

It is important to stress that only the pieces that are
simultaneously $P$-odd and $T$-odd can generate EDM. As the Lorentz-violating
couplings presented in Tables \ref{TableI} are $CPT$-odd, the simultaneously $P$-odd and $T$-odd
pieces turn out to be $CP$-even, but able to generate EDM. Effectively, we
are interested only in the unsuppressed $P$-odd and $T$-odd couplings, that is
\begin{align}
(k_{SV})_{i}(\bar{N}N)(\bar{\psi}\gamma^{i}\psi)\mathbf{,\ }(k_{VP})_{0}(\bar{N}\gamma^{0}%
N)(\bar{\psi}i\gamma_{5}\psi)\ ,
\end{align}
where the factor of $i$ was inserted in order to guarantee the
hermicity. The $CPT$-odd effective Lagrangian, for the possible couplings involving a
rank-$1$ LV tensor, is then%
\begin{align}
\mathcal{L}_{\text{LV}-1}  &  =\sum_{N}\left[ (k_{SV})_{i}(\bar
{N}N)(\bar{\psi}\gamma^{i}\psi)+(k_{VP})_{0}(\bar{N}\gamma^{0}N)(\bar{\psi}%
i\gamma_{5}\psi)
 \right]  \ .\label{Lrank1}%
\end{align}
\begin{table}[ht]
\begin{centering}
\begin{tabular}{|c|c|c|c|c|}
\hline
Coupling &  & $P$-odd and $T$-odd piece & NRL & EDM\tabularnewline
\hline
\hline
 \multicolumn{5}{|l|}{Rank-$2$}\tabularnewline
\hline
\hline
$\left(k_{VV}\right)_{\mu\nu}\left(\bar{N}\gamma^{\mu}N\right)\left(\bar{\psi}\gamma^{\nu}\psi\right)$ &  & \makecell{$\left(k_{VV}\right)_{i0}\left(\bar{N}\gamma^{i}N\right)\left(\bar{\psi}\gamma^{0}\psi\right)$ \\
$\left(k_{VV}\right)_{0i}\left(\bar{N}\gamma^{0}N\right)\left(\bar{\psi}\gamma^{i}\psi\right)$} & \makecell{S \\ NS} & \makecell{-- \\ yes}\tabularnewline
\hline
$\left(k_{AV}\right)_{\mu\nu}\left(\bar{N}\gamma^{\mu}\gamma_{5}N\right)\left(\bar{\psi}\gamma^{\nu}\psi\right)$ &  & none & -- & --\tabularnewline
\hline
$\left(k_{VA}\right)_{\mu\nu}\left(\bar{N}\gamma^{\mu}N\right)\left(\bar{\psi}\gamma^{\nu}\gamma_{5}\psi\right)$ &  & none & -- & --\tabularnewline
\hline
$\left(k_{AA}\right)_{\mu\nu}\left(\bar{N}\gamma^{\mu}\gamma_{5}N\right)\left(\bar{\psi}\gamma^{\nu}\gamma_{5}\psi\right)$ &  & \makecell{$\left(k_{AA}\right)_{0i}\left(\bar{N}\gamma^{0}\gamma_{5}N\right)\left(\bar{\psi}\gamma^{i}\gamma_{5}\psi\right)$\,
\\$\left(k_{AA}\right)_{i0}\left(\bar{N}\gamma^{i}\gamma_{5}N\right)\left(\bar{\psi}\gamma^{0}\gamma_{5}\psi\right)$} & \makecell{S \\ NS} & \makecell{-- \\ yes} \tabularnewline
\hline
$\left(k_{TS}\right)_{\mu\nu}\left(\bar{N}\sigma^{\mu\nu}N\right)\left(\bar{\psi}\psi\right)$ &  & none & -- & --\tabularnewline
\hline
$\left(k_{TP}\right)_{\mu\nu}\left(\bar{N}\sigma^{\mu\nu}N\right)\left(\bar{\psi}i\gamma_{5}\psi\right)$ &  & none & -- & --\tabularnewline
\hline
$\left(k_{ST}\right)_{\mu\nu}\left(\bar{N}N\right)\left(\bar{\psi}\sigma^{\mu\nu}\psi\right)$ &  & none & -- & --\tabularnewline
\hline
$\left(k_{PT}\right)_{\mu\nu}\left(\bar{N}i\gamma_{5}N\right)\left(\bar{\psi}\sigma^{\mu\nu}\psi\right)$ &  & none & -- & --\tabularnewline

\hline
\hline
 \multicolumn{5}{|l|}{Rank-$3$}\tabularnewline
 \hline
\hline 
		$\left(k_{VT}\right)_{\alpha\mu\nu}(\bar{N}\gamma^{\alpha}N)(\bar{\psi}\sigma^{\mu\nu}\psi)$ &  & none & -- & --\tabularnewline
		\hline 
		$\left(k_{AT}\right)_{\alpha\mu\nu}(\bar{N}\gamma^{\alpha}\gamma_{5}N)(\bar{\psi}\sigma^{\mu\nu}\psi)$ &  & \makecell{$\left(k_{AT}\right)_{0ij}(\bar{N}\gamma^{0}\gamma_{5}N)(\bar{\psi}\sigma^{ij}\psi)$
			\\
			$\left(k_{AT}\right)_{i0j}(\bar{N}\gamma^{i}\gamma_{5}N)(\bar{\psi}\sigma^{0j}\psi)$
			\\
			$\left(k_{AT}\right)_{ij0}(\bar{N}\gamma^{i}\gamma_{5}N)(\bar{\psi}\sigma^{j0}\psi)$} & \makecell{S \\ NS \\ NS} & \makecell{-- \\ yes \\ yes}\tabularnewline
	
		\hline 
		$\left(k_{TV}\right)_{\alpha\mu\nu}(\bar{N}\sigma^{\mu\nu}N)(\bar{\psi}\gamma^{\alpha}\psi)$ &  & none & -- & --\tabularnewline
	
		\hline 
		$\left(k_{TA}\right)_{\alpha\mu\nu}(\bar{N}\sigma^{\mu\nu}N)(\bar{\psi}\gamma^{\alpha}\gamma_{5}\psi)$ &  & \makecell{$\left(k_{TA}\right)_{0ij}(\bar{N}\sigma^{ij}N)(\bar{\psi}\gamma^{0}\gamma_{5}\psi)$
			\\
			$\left(k_{TA}\right)_{i0j}(\bar{N}\sigma^{0j}N)(\bar{\psi}\gamma^{i}\gamma_{5}\psi)$
			\\
			$\left(k_{TA}\right)_{ij0}(\bar{N}\sigma^{j0}N)(\bar{\psi}\gamma^{i}\gamma_{5}\psi)$} & \makecell{NS \\ S \\ S} & \makecell{yes \\ -- \\ --}\tabularnewline

\hline
\hline
 \multicolumn{5}{|l|}{Rank-$4$}\tabularnewline
 \hline
\hline 
$\left(k_{TT}\right)_{\alpha\beta\mu\nu}(\bar{N}\sigma^{\alpha\beta}N)(\bar{\psi}\sigma^{\mu\nu}\psi)$ &  & \makecell{$\left(k_{TT}\right)_{0ijk}(\bar{N}\sigma^{0i}N)(\bar{\psi}\sigma^{jk}\psi)$
			\\
			$\left(k_{TT}\right)_{i0jk}(\bar{N}\sigma^{i0}N)(\bar{\psi}\sigma^{jk}\psi)$
			\\
			$\left(k_{TT}\right)_{ij0k}(\bar{N}\sigma^{ij}N)(\bar{\psi}\sigma^{0k}\psi)$
			\\
			$\left(k_{TT}\right)_{ijk0}(\bar{N}\sigma^{ij}N)(\bar{\psi}\sigma^{k0}\psi)$} & \makecell{S \\ S \\ NS \\ NS} & \makecell{-- \\ -- \\ yes \\ yes} \tabularnewline
		
		\hline

\end{tabular}
\par\end{centering}
\caption{General couplings with LV tensors of ranks $2$, $3$ and $4$, and Dirac bilinears. Again, NRL stands for the nonrelativistic limit for the nucleons. Also, in this limit, ``S'' and ``NS'' stand for suppressed and not suppressed, respectively}\label{Table4}
\end{table}

Following the rank-$1$ case, it is interesting to analyze now the %
$e$-$N$ couplings composed by a rank-$2$ LV tensor,
$(k_{XX})_{\mu\nu}$, which are, obviously, $CPT$-even. All
possibilities are listed in Table \ref{Table4}, which contains two $P$-odd
and $T$-odd $e$-$N$ couplings that are not suppressed in the nonrelativistic
limit for the nucleons. These couplings are $CP$-odd, as the usual
Lorentz-preserving EDM terms. Again, similarly as in Eq. (\ref{Redundancy1}), couplings with a rank-$2$ tensor presenting index contractions between the bilinears can be read as particular cases of the rank-$4$ generalization, presented in Table \ref{Table4}.
That said, the $CPT$-even effective Lagrangian is%
\begin{align}
\mathcal{L}_{\text{LV}-2} &  =\sum_{N}\left[  \left(k_{VV}\right)_{0i}\left(
\bar{N}\gamma^{0}N\right)  \left(  \bar{\psi}\gamma^{i}\psi\right)  + \left(k_{AA}\right)_{i0}\left(  \bar{N}\gamma^{i}\gamma_{5}N\right)  \left(  \bar{\psi}%
\gamma^{0}\gamma_{5}\psi\right)  
 \right]
\ .\label{Lrank2}%
\end{align}

After applying the nonrelativistic limit for the nucleons, using the
definitions (\ref{Defsgamma}), the nucleon bilinears yield%
\begin{align}
\bar{N}N\ ,\ \bar{N}\gamma^{0}N\  &  \rightarrow\ n(\boldsymbol{r}),\nonumber\\
\bar{N}\gamma_{i}\gamma_{5}N\ ,\ \bar{N}i\sigma^{0i}\gamma_{5}N\ &
\rightarrow\ -\langle\sigma^{i}\rangle_{N}\cdot n(\boldsymbol{r}),\nonumber \\
\bar{N}\sigma^{ij}N\  &  \rightarrow\ \epsilon_{ijk}\langle\sigma^{k}%
\rangle_{N}\cdot n(\boldsymbol{r})\ ,\label{nonrelNucleon}
\end{align}
where $n(r)$ is the nucleon density, being the same for protons and
neutrons, while $\langle\sigma^{k}\rangle_{N}$ is the effective average spin state of
the nucleon. Notice that, while the densities add coherently to
$A\cdot n(r)$, with $A$ being the atomic mass, the spins do not, so
that only the (unpaired) valence nucleon will count. We can read from the Lagrangians
(\ref{Lrank1}) and (\ref{Lrank2}) the possible EDM contributions
via atomic parity nonconservation methods, which will be illustrated in
Sec. \ref{section3}. For now, we can rewrite the effective Lagrangians (\ref{Lrank1})
and (\ref{Lrank2}) for the valence electron as follows:%
\begin{align}
\mathcal{L}_{\text{LVe}-1}  &  =\left[ \left(k_{SV}\right)_{i}A(\bar{\psi}\gamma
^{i}\psi)+\left(k_{VP}\right)_{0}A(\bar{\psi}i\gamma_{5}\psi)
\right] n(\boldsymbol{r})\ ,\label{Lerank1}%
\end{align}
and
\begin{align}
\mathcal{L}_{\text{LVe}-2}  &  =\left[ A\left(k_{VV}\right)_{0i}\left(
\bar{\psi}\gamma^{i}\psi\right) + \left(k_{AA}\right)_{i0}\langle\sigma^{i}%
\rangle_{N}\left( \bar{\psi}\gamma^{0}\gamma_{5}\psi\right) 
\right] n(\boldsymbol{r}%
)\ .\label{Lerank2}%
\end{align}

Beyond the couplings with rank-$1$ and rank-$2$ LV tensors, we can explore possible interactions involving rank-$3$ and rank-$4$ tensors, which are displayed in Table \ref{Table4}. The $P$- and $T$-odd pieces are shown, with their suppressed or unsuppressed parts. The rank-$3$ couplings of Table \ref{Table4} are $CPT$-odd, so that the highlighted $P$-odd and $T$-odd pieces are $CP$-even. 

Besides the structures presented in Table \ref{Table4}, we could also propose couplings of the form:
\begin{equation}
\left( k_{SX}\right) _{\alpha \mu \nu }(\bar{N}N)(\bar{\psi}\gamma ^{\alpha
}\sigma ^{\mu \nu }\psi)\ . \label{C9}%
\end{equation}%

Nevertheless, using the identity, 
\begin{equation}
\gamma ^{\alpha }\sigma ^{\mu \nu }=-i(g^{\alpha \nu }\gamma ^{\mu }-g^{\mu
	\alpha }\gamma ^{\nu })+\epsilon ^{\beta \alpha \mu \nu }\gamma _{\beta
}\gamma _{5}\ ,
\end{equation}
the coupling (\ref{C9}) is reduced to structures as
\begin{equation}
\left( k_{SV}\right) _{\alpha \mu }^{\text{ \ \ \ }\mu }(\bar{N}N)(\bar{\psi}%
\gamma ^{\alpha }\psi) \, , \, \text{ }\left( k_{SA}\right) _{\alpha \mu }^{\text{ \ \
		\ }\mu }(\bar{N}N)(\bar{\psi}\gamma ^{\alpha }\gamma _{5}\psi)\ ,
\end{equation}%
which are equivalent to rank-$1$ couplings contained in Table \ref{TableI}. Hence, the coupling (\ref{C9}) and its variations, 
\begin{eqnarray}
&&\left( k_{SX}\right) _{\alpha \mu \nu }(\bar{N}N)(\bar{\psi}\gamma ^{\alpha
}\sigma ^{\mu \nu }\psi),\left( k_{SX}\right) _{\alpha \mu \nu }(\bar{N}N)(\bar{\psi}\gamma ^{\alpha }\sigma ^{\mu \nu }\gamma _{5}\psi)\ , \\
&&\left( k_{SX}\right) _{\alpha \mu \nu }(\bar{N}\gamma _{5}N)(\bar{\psi}\gamma
^{\alpha }\sigma ^{\mu \nu }\psi),\left( k_{SX}\right) _{\alpha \mu \nu }(\bar{N%
}\gamma _{5}N)(\bar{\psi}\gamma ^{\alpha }\sigma ^{\mu \nu }\gamma _{5}\psi)\ ,
\end{eqnarray}%
will not be considered.

The couplings of Table \ref{Table4} will also be examined concerning the associated Hamiltonian and spectrum in the next section. In order to accomplish this, 
as already done for the rank-$1$ and rank-$2$ cases, we write the LV Lagrangian consisting of rank-$3$ and rank-$4$ tensor couplings: 
\begin{align}
\mathcal{L}_{\text{LV}-3} &  =\sum_{N}\left[  \left(k_{AT}\right)_{i0j}(\bar{N}\gamma^{i}\gamma_{5}N)(\bar{\psi}\sigma^{0j}\psi)  + 	\left(k_{AT}\right)_{ij0}(\bar{N}\gamma^{i}\gamma_{5}N)(\bar{\psi}\sigma^{j0}\psi) + \left(k_{TA}\right)_{0ij}(\bar{N}\sigma^{ij}N)(\bar{\psi}\gamma^{0}\gamma_{5}\psi) \right],\label{Lrank3}%
\end{align}
\begin{align}
\mathcal{L}_{\text{LV}-4} &  =\sum_{N}\left[  	\left(K_{TT}\right)_{ij0k}(\bar{N}\sigma^{ij}N)(\bar{\psi}\sigma^{0k}\psi)
+\left(K_{TT}\right)_{ijk0}(\bar{N}\sigma^{ij}N)(\bar{\psi}\sigma^{k0}\psi)  \right]
,\label{Lrank4}%
\end{align}
whose nonrelativistic limit for the nucleons, from the definitions (\ref{nonrelNucleon}), imply the following terms for the electron Lagrangian:
\begin{align}
\mathcal{L}_{\text{LVe}-3} &  =\sum_{N}\left[  \left(k_{AT}\right)_{i0j}(-\langle\sigma^{i}\rangle_{N})(\bar{\psi}\sigma^{0j}\psi)  + 	\left(k_{AT}\right)_{ij0}(-\langle\sigma^{i}\rangle_{N})(\bar{\psi}\sigma^{j0}\psi) + \left(k_{TA}\right)_{0ij}(\epsilon_{ijk}\langle\sigma^{k}%
\rangle_{N})(\bar{\psi}\gamma^{0}\gamma_{5}\psi) \right]\cdot n(\boldsymbol{r})
\ ,\label{Lrank3B}%
\end{align}
and
\begin{align}
\mathcal{L}_{\text{LVe}-4} &  =\sum_{N}\left[  	\left(k_{TT}\right)_{ij0k}(\epsilon_{ijl}\langle\sigma^{l}%
\rangle_{N})(\bar{\psi}\sigma^{0k}\psi)
+\left(k_{TT}\right)_{ijk0}(\epsilon_{ijl}\langle\sigma^{l}%
\rangle_{N})(\bar{\psi}\sigma^{k0}\psi)  \right]\cdot n(\boldsymbol{r})\ 
.\label{Lrank4B}%
\end{align}

Below we will show how to read the EDM contributions from a $P$-odd and $T$-odd Hamiltonian piece.

\section{Spectrum Shifts and EDMs}\label{section3}

In order to illustrate how $P$-odd and $T$-odd $e$-$N$ couplings generate EDM,
we follow Refs. \cite{AtMolPhys, LeptonEDM, Sandars2}. Consider the
relativistic Hamiltonian of the valence electron under a radial potential,
$\Phi_{\text{int}}(\boldsymbol{r}),$
\begin{equation}
H_{0}=\boldsymbol{\alpha}\cdot\boldsymbol{p}+m\gamma^{0}-e\Phi_{\text{int}%
}(\boldsymbol{r}),\label{H0}%
\end{equation}
whose eigenstates are $|\psi_{0}\rangle$. If an external electric field,
$\boldsymbol{E}=E_{z}\hat{z}$, is applied, the solutions, to first order in
$\boldsymbol{E}$, become
\begin{equation}
|\Psi\rangle=|\psi_{0}\rangle+eE_{z}|\eta\rangle\ ,
\end{equation}
where
\begin{equation}
|\eta\rangle=\sum_{n\neq0}\frac{|\psi_{n}\rangle\langle\psi_{n}|z|\psi
_{0}\rangle}{E_{0}-E_{n}}\ .
\end{equation}

We will use the data available in Ref. \cite{AtMolPhys} for the thallium
($A=205$ and $Z=81$) valence electron under a modified Tiez potential,
$\Phi_{\text{int}}(\boldsymbol{r})$. The states $|\psi_{0}\rangle$ and
$|\eta\rangle$ have opposite parity (the ground state $|\psi_{0}\rangle$ has
$l=1$ and $|\eta\rangle$ has $l=0$). The expressions for the unperturbed
wave function $\psi_0$ (with $l=1$) and for the state $\eta$ (with $l=0)$ are, respectively,%
\begin{equation}
\left(\psi_0\right)_{J=\frac{1}{2},m=\frac{1}{2}}^{l}=\left(
\begin{array}
[c]{c}%
\frac{i}{r}G_{l,J=\frac{1}{2}}(r)\phi_{\frac{1}{2},\frac{1}{2}}^{l}\\
\frac{1}{r}F_{l,J=\frac{1}{2}}(r)\left(  \boldsymbol{\sigma}\cdot
\hat{\boldsymbol{r}}\right)  \phi_{\frac{1}{2},\frac{1}{2}}^{l}%
\end{array}
\right)  ,\label{psi}%
\end{equation}
\begin{equation}
\eta_{J=\frac{1}{2},m=\frac{1}{2}}^{l}=\left(
\begin{array}
[c]{c}%
\frac{i}{r}G_{l,J=\frac{1}{2}}^{S}(r)\phi_{\frac{1}{2},\frac{1}{2}}^{l}\\
\frac{1}{r}\left(  F_{l,J=\frac{1}{2}}^{S}(r)\boldsymbol{\sigma}\cdot
\hat{\boldsymbol{r}}\right)  \phi_{\frac{1}{2},\frac{1}{2}}^{l}%
\end{array}
\right)  \ ,\label{eta}%
\end{equation}
where,%
\begin{equation}
\phi_{\frac{1}{2},\frac{1}{2}}^{l=0}=\left(
\begin{array}
[c]{c}%
Y_{0}^{0}\\
0
\end{array}
\right)  \ ,\ \ \ \phi_{\frac{1}{2},\frac{1}{2}}^{l=1}=\left(
\begin{array}
[c]{c}%
\sqrt{\frac{1}{3}}Y_{1}^{0}\\
-\sqrt{\frac{2}{3}}Y_{1}^{1}%
\end{array}
\right)  \ ,
\end{equation}
where $Y_{l}^{m}$ corresponds to the normalized spherical harmonics.
The superscript $S$ stands for Sternheimer solutions,
which hold for the Sternheimer equation, $\left(  H_{0}%
-E_{0}\right)  |\eta\rangle=z|\psi_{0}\rangle$ - see Refs.
\cite{Sternheimer1,Sternheimer2}. While the unperturbed spinor is normalized
to unity, $\eta_{J=\frac{1}{2},m=\frac{1}{2}}^{l}$ is not. It is nevertheless required that
$G_{l,J=\frac{1}{2}}^{S}(r)\,,\,F_{l,J=\frac{1}{2}}^{S}(r)\rightarrow0$ far
from the origin. These solutions have been used, for instance, in
evaluating enhancement factors of atomic EDMs and atomic polarizations -- see
\cite{LeptonEDM,AtMolPhys,AtPol}.

In order to calculate the EDMs arising from the $e$-$N$ couplings, we start by extracting
from the Lagrangian (\ref{Lerank1}), related to the rank-$1$ LV tensor, the following Hamiltonian contributions
\begin{align}
H_{\text{LV}e-1} &  =\left[-\left(k_{SV}\right)_{i}A\gamma^{0}\gamma^{i}
-\left(k_{VP}\right)_{0}Ai\gamma^{0}\gamma_{5}
\right]
n(\boldsymbol{r})\ .\label{LVHrank1}%
\end{align}
The corresponding energy shift for a given $P$-odd and $T$-odd
Hamiltonian piece $H_{P,T}$ is
\begin{equation}
\Delta E=\langle\Psi|H_{P,T}|\Psi\rangle=2eE_{z}\Re\left[  \int\psi
_{0}^{\dagger}H_{P,T}\eta d^{3}r\right]. \label{Energy}%
\end{equation}
where we used that $|\Psi\rangle=|\psi_{0}\rangle+eE_{z}|\eta\rangle$,
remembering that $H_{P,T}$ is $P$-odd. From Eq.
(\ref{Energy}), the EDM magnitude is read as the factor multiplying the electric field on the right hand side, $|d_{\text{equiv}}|=2e\Re\left[
\langle\psi_{0}|H_{P,T}|\eta\rangle\right]  $, that is,
\begin{equation}
\frac{\Delta E}{E_{z}}=2e\Re\left[  \langle\psi_{0}|H_{P,T}|\eta
\rangle\right]  \equiv d_{\text{equiv}}\ .\label{EDMequiv}%
\end{equation}

In this case, $d_{\text{equiv}}$ is an EDM-equivalent due to
the $P$-odd and $T$-odd LV interactions. Applying this prescription to the
Hamiltonian (\ref{LVHrank1}) and using the spinors (\ref{psi}) and
(\ref{eta}), we notice that the contribution from $\left(k_{SV}\right)_{i}$ has no real part, so that it does not yield $d_{\text{equiv}}$, according to Eq.
(\ref{EDMequiv}) and Eqs. (\ref{expalpha}-\ref{expigamma}). As for the term $\left(k_{VP}\right)_{0}$, its contribution is similar to
the one arising from the scalar-pseudoscalar coupling shown in Eq. (\ref{UsualCase}), that is
\begin{equation}
|d^{(1)}_{1-\text{equiv}}|=2eA\left(k_{VP}\right)_{0}n(r)|\Re\left[  \langle\psi_{0}|\left(
i\gamma^{0}\gamma_{5}\right)  |\eta\rangle\right]  |\ ,
\end{equation}
which can be written as
\begin{equation}
|d^{(1)}_{1-\text{equiv}}|=2e\left(k_{VP}\right)_{0}\frac{3A}{4\pi R_{\text{Nucleus}%
}^{3}}\int_{0}^{R_{\text{Nucleus}}}\left[  F^{S}(r)G(r)+G^{S}(r)F(r)\right]
dr\ , \label{d1}
\end{equation}
where we used a uniform nucleon density, $n=3/4\pi R_{\text{Nucleus}%
}^{3}$.
Supposing $P$-odd and $T$-odd $e$-$N$ couplings are the only source of atomic
EDM, we can set bounds on $\left(k_{VP}\right)_{0}$ by using the data of the thallium
atom, available in Ref. \cite{AtMolPhys,PNC}, and the current upper limit on
the electron's EDM, $|\mathbf{d}_{e}|<1.1\times10^{-29}\,e\cdot\text{cm}$ in
Ref. \cite{EDMnature2018}. This implies the following upper bound%
\begin{equation}
|\left(k_{VP}\right)_{0}|<1.6\times10^{-15}\,(\text{GeV})^{-2}\ .\label{Bound1}%
\end{equation}

Since the limit (\ref{Bound1}) holds for the zero component of the vector, it is not subjected
to sidereal variations, unlike the ones attained in Refs. \cite{Jonas1,Jonas2}. We must point out, however, that this term, being $CPT$-odd, also violates $C$, so that, unlike usual EDMs, it may play a role in the baryon asymmetry in the universe. 

As for the Hamiltonian contribution arising from couplings with a rank-$2$ LV tensor
\begin{align}
H_{\text{LVe}-2}  &  =\left[ -A\left(k_{VV}\right)_{0i}\gamma^0\gamma^{i} - \left(k_{AA}\right)_{i0}\langle\sigma^{i}
\rangle_{N}\gamma_{5}\right] n(\boldsymbol{r}%
)\ ,\label{Herank2}%
\end{align}
we find that its pieces do not yield $d_{\text{equiv}}$, according to Eq. (\ref{EDMequiv}), for they generate imaginary contributions. In summary, for internal products of the form $\langle\psi_{0}| \mathcal{O} |\eta\rangle$, in which the operator $\mathcal{O}$ is a $4\times4$ matrix, the following results are useful:
\begin{align}
\langle\psi_{0}| \gamma^0\gamma^{i} |\eta\rangle & =i\delta_{i3}\int_{0}^{R_{\text{Nucleus}}}\left[  \frac{1}{3}F^{S}(r)G(r)+G^{S}(r)F(r)\right]dr\ \label{expalpha}, \\
\langle\psi_{0}| \gamma_5 |\eta\rangle &=i\int_{0}^{R_{\text{Nucleus}}}\left[-F^{S}(r)G(r)+G^{S}(r)F(r)\right]dr\  \label{expgamma5}, \\
\langle\psi_{0}| i \gamma^{i} |\eta\rangle & =\delta_{i3}\int_{0}^{R_{\text{Nucleus}}}\left[  -\frac{1}{3}F^{S}(r)G(r)+G^{S}(r)F(r)\right]dr\  \label{expigamma},
\end{align}
where the $\delta_{i3}$ means that only the $i=3$ component survives, due to the structure of the $\gamma^3$ matrix, while the analogue contributions for $i=1,2$ vanish.

For the rank-$3$ and rank-$4$ cases, the Hamiltonians are
\begin{equation}
H_{\text{LVe}-3}=\left[\left(k_{AT}\right)_{i0j}\langle\sigma^{i}\rangle_{N}i\gamma^{0}\alpha^{j}-\left(k_{AT}\right)_{ij0}\langle\sigma^{i}\rangle_{N}i\gamma^{0}\alpha^{j}-\left(k_{TA}\right)_{0ij}\epsilon_{ijk}\langle\sigma^{k}\rangle_{N}\gamma_{5}\right]\cdot n(\boldsymbol{r})\ , \label{Herank3}
\end{equation}
\begin{equation}
H_{\text{LVe}-4}=\left[-\left(k_{TT}\right)_{ij0k}\epsilon_{ijl}\langle\sigma^{l}\rangle_{N}i\gamma^{0}\alpha^{k}+\left(k_{TT}\right)_{ijk0}\epsilon_{ijl}\langle\sigma^{l}\rangle_{N}i\gamma^{0}\alpha^{k}\right]\cdot n(\boldsymbol{r})\ ,\label{Herank4}
\end{equation}
in which the term $\left(k_{TA}\right)_{0ij}$, according to Eq. (\ref{expgamma5}), yields no real contribution. Concerning the
remaining pieces, the Hamiltonian (\ref{Herank4}) can also be read as
\begin{equation}
H_{\text{LVe}-4}=\left[-\left(K_{TT}\right)_{0kl}\langle\sigma^{l}\rangle_{N}i\gamma^{k}+\left(K_{TT}\right)_{0kl}\langle\sigma^{l}\rangle_{N}i\gamma^{k}\right]\cdot n(\boldsymbol{r})\ ,\label{Hrank4}
\end{equation}
with the redefinition $\left(k_{TT}\right)_{ij0k}\epsilon_{ijl}=\left(K_{TT}\right)_{0kl}$.  The EDM contributions from the rank-$3$ Hamiltonian are
\begin{align}
d_{3-\text{equiv}}^{(1)} & =2e\left(k_{AT}\right)_{i03}\langle\sigma^{i}\rangle_{N}\Re\left[\langle\psi_{0}|i\gamma^{3}|\eta\rangle\right]\ \label{d2},\\ 
d_{3-\text{equiv}}^{(2)} & =-2e\left(k_{AT}\right)_{i30}\langle\sigma^{i}\rangle_{N}\Re\left[\langle\psi_{0}|i\gamma^{3}|\eta\rangle\right]\ ,\label{d3}
\end{align}
and from rank-$4$,
\begin{align}
d_{4-\text{equiv}}^{(1)} & =-2e\left(K_{TT}\right)_{03l}\langle\sigma^{l}\rangle_{N}\Re\left[\langle\psi_{0}|i\gamma^{3}|\eta\rangle\right]\ \label{d4},\\ 
d_{4-\text{equiv}}^{(2)} & =2e\left(K_{TT}\right)_{30l}\langle\sigma^{l}\rangle_{N}\Re\left[\langle\psi_{0}|i\gamma^{3}|\eta\rangle\right]\ .\label{d5}
\end{align}

We point out that the presence of the factor $\epsilon_{ijk}$ in Eq. (\ref{Herank4}) induces an effect similar to a rotation on the nucleon's spin $\langle\sigma^{l}\rangle_{N}$ or background vector, coupling orthogonal components of $\langle
\sigma^{l}\rangle_{N}$ and $\left(k_{TT}\right)_{ij03}$. Accordingly, say we pick $i=1$, then $j=2$, and we must use a nucleon in the spin state $\langle\sigma^{z}\rangle_{N}$ in order to obtain the EDM contribution arising from $\left(k_{TT}\right)_{1203}$, which corresponds to constrain $\left(K_{TT}\right)_{033}$.

Assuming that the thallium valence nucleon has spin $\langle
\sigma^{z}\rangle_{N}=\pm1$ \cite{PNC}, and that the atomic integral in Eq. (\ref{expigamma}) has the same magnitude as the one in
Eq. (\ref{d1}), we attain, for the rank-$3$ couplings, the upper bounds
\begin{align}
|\left(k_{AT}\right)_{i03}|  & <3.2\times10^{-13}\,(\text{GeV})^{-2}\ \label{bd2},\\
|\left(k_{AT}\right)_{i30}|  & <3.2\times10^{-13}\,(\text{GeV}\label{bd3}
)^{-2}\ ,
\end{align}
and for the rank-$4$:
\begin{align}
|\left(K_{TT}\right)_{03l}|  & <3.2\times10^{-13}\,(\text{GeV})^{-2}\ \label{bd4}, \\
|\left(K_{TT}\right)_{30l}|  & <3.2\times10^{-13}\,(\text{GeV})^{-2}\ \label{bd5}.
\end{align}

In these cases, the bounds suffer sidereal variations,  
for these are in fact measured in the Lab's reference frame, in which these tensor components are not constant. This issue will be addressed in the Sec. \ref{SiderealAnalysis} while the constraints are summarized in Table \ref{TableBounds}.

\subsection{Sidereal analysis}\label{SiderealAnalysis}

Because the LV background tensors are constant only in an inertial reference frame (RF), such as the Sun's rest frame,
it is necessary, therefore, to show how to translate
these bounds to the Earth-located Lab's RF, at the colatitude $\chi$, rotating
around the Earth's axis with angular velocity $\Omega$. In short, the bounds should be written in
 terms of the tensor components in the Sun's RF. For experiments up to
a few weeks long, the transformation law for a rank-$2$ tensor, say $B_{\mu\nu}$,
according to Refs. \cite{Jonas1,Sidereal}, is merely a spatial rotation,%
\begin{align}
B_{\mu\nu}^{\text{(Lab)}}& =\mathcal{R}_{\mu\alpha}\mathcal{R}_{\nu\beta
}B_{\alpha\beta}^{\text{(Sun)}},\label{rotationLaw2}
\end{align}
where
\begin{equation}
R_{\mu\nu}=%
\begin{pmatrix}
1 & 0 & 0 & 0\\
0 & \cos\chi\cos\Omega t & \cos\chi\sin\Omega t & -\sin\chi\\
0 & -\sin\Omega t & \cos\Omega t & 0\\
0 & \sin\chi\cos\Omega t & \sin\chi\sin\Omega t & \cos\chi \label{Rotation}%
\end{pmatrix}
,
\end{equation}
in which the first line and column were included for completeness. According to the transformation law
 (\ref{rotationLaw2}), the components $\left(k_{AT}\right)_{i03}$ and $\left(k_{AT}\right)_{i30}$ transform as
\begin{equation}
\left(k_{AT}\right)^{\text{(Lab)}}_{i03}=
\mathcal{R}_{ik}\mathcal{R}_{3l}\left(k_{AT}\right)^{\text{(Sun)}}_{k0l}\ ,
\end{equation}
\begin{equation}
\left(k_{AT}\right)^{\text{(Lab)}}_{i30}=
 \mathcal{R}_{ik}\mathcal{R}_{3l}\left(k_{AT}\right)^{\text{(Sun)}}_{kl0}\ ,
\end{equation}
whose time averages for $l=1,2,3$ yield, respectively:
\begin{align}
\langle \left(k_{AT}\right)^{\text{(Lab)}}_{103}\rangle & =
\frac{1}{4}\left[\left(k_{AT}\right)^{\text{(Sun)}}_{101}+\left(k_{AT}\right)^{\text{(Sun)}}_{202}-2\left(k_{AT}\right)^{\text{(Sun)}}_{303}\right]\sin2\chi \nonumber \\
\langle \left(k_{AT}\right)^{\text{(Lab)}}_{203}\rangle & =
\left[-\left(k_{AT}\right)^{\text{(Sun)}}_{102}+\left(k_{AT}\right)^{\text{(Sun)}}_{201}\right]\sin\chi \nonumber \\
\langle \left(k_{AT}\right)^{\text{(Lab)}}_{303}\rangle & =
\left[\frac{1}{2}\left( \left(k_{AT}\right)^{\text{(Sun)}}_{101}+\left(k_{AT}\right)^{\text{(Sun)}}_{202} \right)\sin^2\chi+ \left(k_{AT}\right)^{\text{(Sun)}}_{303}\cos^2\chi \right]\ ,
\end{align}
and similar expressions for $\left(k_{AT}\right)^{\text{(Lab)}}_{130}$, $\left(k_{AT}\right)^{\text{(Lab)}}_{230}$ and $\left(k_{AT}\right)^{\text{(Lab)}}_{330}$. The same procedure can be applied to the components of $\left(K_{TT}\right)_{03l}$ for $l=1,2,3$, yielding, respectively, 
\begin{align}
\langle \left(K_{TT}\right)^{\text{(Lab)}}_{031}\rangle & =
\frac{1}{4}\left[\left(K_{TT}\right)^{\text{(Sun)}}_{011}+\left(K_{TT}\right)^{\text{(Sun)}}_{022}-2\left(K_{TT}\right)^{\text{(Sun)}}_{033}\right]\sin2\chi \nonumber \\
\langle \left(K_{TT}\right)^{\text{(Lab)}}_{032}\rangle & =
\left[-\left(K_{TT}\right)^{\text{(Sun)}}_{012}+\left(K_{TT}\right)^{\text{(Sun)}}_{201}\right]\sin\chi \nonumber \\
\langle \left(K_{TT}\right)^{\text{(Lab)}}_{303}\rangle & =
\left[\frac{1}{2}\left( \left(K_{TT}\right)^{\text{(Sun)}}_{101}+\left(K_{TT}\right)^{\text{(Sun)}}_{202} \right)\sin^2\chi+ \left(K_{TT}\right)^{\text{(Sun)}}_{303}\cos^2\chi \right]\ ,
\end{align}
and similar expressions for the components  $\left(K_{TT}\right)_{30l}$. Finally, the sidereal variations of the bounds (\ref{bd4}) are displayed as:%
\begin{align}
|\frac{1}{4}\left[\left(K_{TT}\right)^{\text{(Sun)}}_{011}+\left(K_{TT}\right)^{\text{(Sun)}}_{022}-2\left(K_{TT}\right)^{\text{(Sun)}}_{033}\right]\sin 2 \chi |&<3.2\times10^{-13}\,(\text{GeV})^{-2} \nonumber \\
|\left[\left(K_{TT}\right)^{\text{(Sun)}}_{012}-\left(K_{TT}\right)^{\text{(Sun)}}_{021}\right]\sin\chi |&<3.2\times10^{-13}\,(\text{GeV})^{-2}\nonumber \\
|\left[\frac{1}{2}\left( \left(K_{TT}\right)^{\text{(Sun)}}_{011}+\left(K_{TT}\right)^{\text{(Sun)}}_{022} \right)\sin^2\chi+ \left(K_{TT}\right)^{\text{(Sun)}}_{033}\cos^2\chi \right] |&<3.2\times10^{-13}\,(\text{GeV})^{-2}\ ,
\end{align}
For clarity, the bounds (\ref{bd2})-(\ref{bd5}), in terms of the Sun's RF quantities, are listed in the Table \ref{TableBounds} for $l=1,2,3$.
\begin{table}[h]
\begin{centering}
\begin{tabular}{|l|c|}
\hline 
Component & Upper bound\tabularnewline
\hline 
\hline 
$|\left(k_{VP}\right)_{0}^{\text{(Sun)}}|$ & $1.6\times10^{-15}(\text{GeV})^{-2}$\tabularnewline
\hline 
$|\frac{1}{4}\left[\left(k_{AT}\right)^{\text{(Sun)}}_{101}+\left(k_{AT}\right)^{\text{(Sun)}}_{202}-2\left(k_{AT}\right)^{\text{(Sun)}}_{303}\right]\sin2\chi\ |$ & $3.2\times10^{-13}(\text{GeV})^{-2}$\tabularnewline
\hline 
$|\left[-\left(k_{AT}\right)^{\text{(Sun)}}_{102}+\left(k_{AT}\right)^{\text{(Sun)}}_{201}\right]\sin\chi|$ & $3.2\times10^{-13}(\text{GeV})^{-2}$\tabularnewline
\hline 
$|\left[\frac{1}{2}\left( \left(k_{AT}\right)^{\text{(Sun)}}_{101}+\left(k_{AT}\right)^{\text{(Sun)}}_{202} \right)\sin^2\chi+ \left(k_{AT}\right)^{\text{(Sun)}}_{303}\cos^2\chi \right]|$ & $3.2\times10^{-13}(\text{GeV})^{-2}$\tabularnewline
\hline 
$|\frac{1}{4}\left[\left(k_{AT}\right)^{\text{(Sun)}}_{110}+\left(k_{AT}\right)^{\text{(Sun)}}_{220}-2\left(k_{AT}\right)^{\text{(Sun)}}_{330}\right]\sin2\chi |$ & $3.2\times10^{-13}(\text{GeV})^{-2}$\tabularnewline
\hline
$|\left[-\left(k_{AT}\right)^{\text{(Sun)}}_{120}+\left(k_{AT}\right)^{\text{(Sun)}}_{210}\right]\sin\chi |$ & $3.2\times10^{-13}(\text{GeV})^{-2}$\tabularnewline
\hline 
$|\left[\frac{1}{2}\left( \left(k_{AT}\right)^{\text{(Sun)}}_{110}+\left(k_{AT}\right)^{\text{(Sun)}}_{220} \right)\sin^2\chi+ \left(k_{AT}\right)^{\text{(Sun)}}_{330}\cos^2\chi \right] |$ & $3.2\times10^{-13}(\text{GeV})^{-2}$\tabularnewline
\hline
$|\frac{1}{4}\left[\left(K_{TT}\right)^{\text{(Sun)}}_{011}+\left(K_{TT}\right)^{\text{(Sun)}}_{022}-2\left(K_{TT}\right)^{\text{(Sun)}}_{033}\right]\sin 2 \chi |$ & $3.2\times10^{-13}(\text{GeV})^{-2}$\tabularnewline
\hline
$|\left[\left(K_{TT}\right)^{\text{(Sun)}}_{012}-\left(K_{TT}\right)^{\text{(Sun)}}_{021}\right]\sin\chi |$ & $3.2\times10^{-13}(\text{GeV})^{-2}$\tabularnewline
\hline 
$|\left[\frac{1}{2}\left( \left(K_{TT}\right)^{\text{(Sun)}}_{011}+\left(K_{TT}\right)^{\text{(Sun)}}_{022} \right)\sin^2\chi+ \left(K_{TT}\right)^{\text{(Sun)}}_{033}\cos^2\chi \right] |$ & $3.2\times10^{-13}(\text{GeV})^{-2}$\tabularnewline
\hline 
$|\frac{1}{4}\left[\left(K_{TT}\right)^{\text{(Sun)}}_{101}+\left(K_{TT}\right)^{\text{(Sun)}}_{202}-2\left(K_{TT}\right)^{\text{(Sun)}}_{303}\right]\sin 2 \chi |$ & $3.2\times10^{-13}(\text{GeV})^{-2}$\tabularnewline
\hline 
$|\left[\left(K_{TT}\right)^{\text{(Sun)}}_{102}-\left(K_{TT}\right)^{\text{(Sun)}}_{201}\right]\sin\chi |$ & $3.2\times10^{-13}(\text{GeV})^{-2}$\tabularnewline
\hline 
$|\left[\frac{1}{2}\left( \left(K_{TT}\right)^{\text{(Sun)}}_{101}+\left(K_{TT}\right)^{\text{(Sun)}}_{202} \right)\sin^2\chi+ \left(K_{TT}\right)^{\text{(Sun)}}_{303}\cos^2\chi \right] |$ & $3.2\times10^{-13}(\text{GeV})^{-2}$\tabularnewline
\hline 

\end{tabular}
\par\end{centering}
\caption{Bounds on the LV tensors of ranks ranging from $1$ to $4$.}
\label{TableBounds}

\end{table}

\section{Conclusion and final remarks}\label{section4}

One point to be noted is that the $4$-fermion couplings here presented involve two  kinds of fermion spinors, that is, we
consider both $\left(k_{SV}\right)_{\mu}(\bar{N}N)(\bar{\psi}\gamma^{\mu}\psi)$ and $\left(k_{VS}\right)_{\mu
}(\bar{N}\gamma^{\mu}N)(\bar{\psi}\psi).$ Such a distinction is not
important when one does not differ between the interacting fermion bilinears, but it should be taken into account for dimension-$6$ electron-nucleon interactions. Accordingly, the couplings presented in Tables \ref{TableI} and \ref{Table4} are displayed in Ref. \cite{LVArbDim}, the only difference is that here we consider twice as many couplings, since we are dealing with different spinors. To illustrate the redundancies, we have included comments and the couplings in Table \ref{Table2}, which are particular parametrizations of the ones of rank-$3$ listed in Table \ref{Table4}.

Still concerning Ref. \cite{LVArbDim}, it is worth mentioning that
the first line of the dimension-$6$ couplings of its Table III contains %
\begin{equation}
\left(  k_{SS}\right)  (\bar{N}N)(\bar{\psi}\psi),\left(  k_{PP}\right)  (\bar
{N}i\gamma_{5}N)(\bar{\psi}i\gamma_{5}\psi),\left(  k_{SP}\right)  (\bar{N}%
N)(\bar{\psi}i\gamma_{5}\psi),\left(  k_{PS}\right)  (\bar{N}i\gamma_{5}N)(\bar
{\psi}\psi)\ ,
\end{equation}
that do not violate Lorentz symmetry. From those, only the last two are $P$-odd and $T$-odd and generate EDM, and only the third one is not
suppressed, being equivalent to %
\begin{equation}
\mathcal{L}_{\psi\psi}^{(6)}=i\left(  \kappa_{SP}\right)  \left(  \bar{\psi
}\psi\right)  \left(  \bar{\psi}\gamma^{5}\psi\right)  .\label{AlanSPcoupling}%
\end{equation}

Such a coupling is identical to the standard scalar-pseudoscalar
electron-nucleon interaction in Eq. (\ref{UsualCase}) \cite{AtMolPhys,LeptonEDM,SMEDM2}. It yields%
\begin{equation}
d_{\text{equiv}-\kappa_{SP}}=2eA\kappa_{SP}\frac{3}{4\pi R_{\text{Nucleus}%
}^{3}}\int_{0}^{R_{\text{Nucleus}}}\left(  F^{S}(r)G(r)+G^{S}(r)F(r)\right)
dr\ ,
\end{equation}
and must fulfill %
\begin{equation}
\kappa_{SP}<1.6\times10^{-15}\,(\text{GeV})^{-2}\ .\label{BoundKostelecky}
\end{equation}

Clearly, due to the scalar form of the coupling
(\ref{AlanSPcoupling}), the coefficient $\kappa_{SP}$ does not suffer
sidereal variations. 

In outlining the procedure, we stress that the possibilities of couplings with LV tensors of ranks ranging from $1$ to $4$ were listed in Tables \ref{TableI} and \ref{Table4}. Their behavior under $C$, $P$ and $T$ was crucial to extract the components compatible with EDM generation. After applying the nonrelativistic limit for the nucleons, a few candidates were suppressed and no longer considered. The remaining unsuppressed couplings had their Hamiltonian pieces evaluated in Eqs. (\ref{LVHrank1}), (\ref{Herank2}), (\ref{Herank3}), and (\ref{Herank4}), then their respective EDM-equivalent contributions were calculated as in Eq. (\ref{EDMequiv}), via atomic parity nonconservation methods. Also, as mentioned in Sec. \ref{section2}, due to their matrix structure, some couplings yield no real contribution to the energy shift, i.e., null $d_{\text{equiv}}$, being then discarded. The surviving terms had their magnitudes limited using the electron's EDM data, which is justified by the fact that the measured atomic EDM (attributed to either an unpaired electron or nucleon, depending on the experiment) may be due to $P$-odd and $T$-odd $e$-$N$ interactions as well -- thus, we have just chosen the most restrictive upper bound.

The most stringent upper bounds of $1.6\times10^{-15}\,(\text{GeV})^{-2}$, in Eqs. (\ref{Bound1}) and (\ref{BoundKostelecky}), were set on the terms proportional to the atomic mass $A$, i.e. the terms $\left(k_{VP}\right)_{0}$ and $\kappa_{SP}$ -- these are the dominant contributions. The other bounds are about $2$ orders of magnitude less restrictive, due to the fact that the nucleons' spins do not add up coherently. Moreover, the bounds with spatial indices suffer sidereal variations due to the Earth's rotation and Lab's location, and were transformed according to the rule (\ref{rotationLaw2}) and then time averaged, providing the bounds in Table \ref{TableBounds}.


\begin{acknowledgments}
The authors are grateful to CNPq, CAPES, FAPESP and FAPEMA (Brazilian research
agencies) for invaluable financial support. We thank Dr. Nodoka Yamanaka and Dr. Alan Kostelecky for the useful and relevant feedback and private communications. We also thank Dr. Paul Mansfield and Jo\~{a}o A. A. de Simões, for the interesting discussions.
This study was financed in part by
the Coordena\c{c}\~{a}o de Aperfei\c{c}oamento de Pessoal de N\'{\i}vel
Superior--Brazil (CAPES) -- Finance Code 001, and by FCT Portugal ref. UID/FIS/04564/2016. M. S. thanks CNPq [303482/2017-6]. 
M.M.F. is also grateful to FAPEMA/UNIVERSAL/00880/15; FAPEMA/PRONEX 01452-14; CNPq/PRODUTIVIDADE/308933/2015-0.
\end{acknowledgments}


\begin{thebibliography}{99}                                                                                               %


\bibitem {EDM1}W. Bernreuther and M. Suzuki, Rev. Mod. Phys. \textbf{63}, 313 (1991).

\bibitem {EDM2}J.S.M. Ginges and V.V. Flambaum, Phys. Rep. 397, \textbf{397},
63 (2004); Phys. Rev. A 65, 032113 (2002); J. Jesus and J. Engel, Phys. Rev. C
72, 045503 (2005).


\bibitem {EDM3}J. Engel, M.J. Ramsey-Musolf and U. van Kolck, Electric dipole
moments of nucleons, nuclei and atoms: the Standard Model and beyond, Prog.
Part. Nucl. Phys. 71 (2013) 21; T. Chupp and M. Ramsey-Musolf, Electric dipole
moments: a global analysis, Phys. Rev. C 91 (2015) 035502.


\bibitem {LeptonEDM}B. Lee Roberts and W. J. Marciano, \textit{Lepton dipole
Moments}, Advanced Series on Directions in High Energy Physics (Word
Scientific, Singapore, 2010).


\bibitem {Yamanaka}N. Yamanaka, Int. J. Mod. Phys. E 26, 1730002 (2017); N.
Yamanaka, B. Sahoo, N. Yoshinaga, T. Sato, K. Asahi, B. Das, Eur. Phys. J. A 53, 54 (2017).



\bibitem {Pospelov2005}M. Pospelov and A. Ritz, Electric dipole moments as
probes of new physics, Annals of Physics 318, 119 (2005).


\bibitem {Tviolation}J. P. Lees \textit{et al.} (The BABAR Collaboration),
Phys. Rev. Lett. \textbf{109}, 211801(2012).



\bibitem {Sakharov}A. D. Sakharov, Pis'ma Zh. Eksp. Teor. Fiz. \textbf{5}, 32
(1967) [JETP Lett. \textbf{5}, 24 (1967)].


\bibitem {EDMnature2018}ACME Collaboration, Nature \textbf{562}, 355 (2018).

\bibitem {Measure1}R. Engfer and H.K. Walter, Annu. Rev. Nucl. Part. Sci.
\textbf{36}, 327 (1986).

\bibitem {Measure2}B.C. Regan, E.D. Commins, C.J. Schmidt, and D. DeMille,
Phys. Rev. Lett. \textbf{88}, 071805 (2002); J. J. Hudson, D. M. Kara, I. J.
Smallman, B. E. Sauer, M. R. Tarbutt, and E. A. Hinds, Nature \textbf{473},
493 (2011); D. M. Kara, I. J. Smallman, J. J. Hudson, B. E. Sauer, M. R
Tarbutt and E. A. Hinds, New J. Phys. \textbf{14}, 103051 (2012).

\bibitem {Baron}J. Baron \textit{et al}, (ACME Collaboration), Order of
magnitude smaller limit on the electric dipole moment of the electron, Science
\textbf{343}, 269 (2014).


\bibitem {SMEDM1}N. Cabibbo, Phys. Rev. Lett. \textbf{10}, 531 (1963); M.
Pospelov and I. B. Khriplovich, Sov. J. Nucl. Phys. 53, 638 (1991).

\bibitem {SMEDM2}M. Pospelov and A. Ritz, Phys. Rev. D \textbf{89}, 056006 (2014).



\bibitem {Flambaum1}O. P. Sushkov, V. V. Flambaum, and I. B. Khriplovich, Zh.
Eksp. Teor. Fiz. \textbf{87}, 1521 (1984) [Sov. Phys. JETP \textbf{60}, 873
(1984)]; V. V. Flambaum, I. B. Khriplovich, and O. P. Sushkov, Nucl. Phys.
A\textbf{449}, 750 (1986).





\bibitem {Schiff}I. I. Schiff, Phys. Rev. \textbf{132}, 2194 (1963).



\bibitem {Sandars1}P. G. H. Sandars, J. Phys. B: At. Mol. Phys., \textbf{1},
499 (1968).



\bibitem {Sandars2}P. G. H. Sandars, J. Phys. B: At. Mol. Phys. \textbf{1},
511 (1968).



\bibitem {Sandars0}P. G. H. Sandars, Phys. Lett. \textbf{14}, 194 (1965).



\bibitem {Sternheimer2}P. G. H. Sandars and R. M. Sternheimer, Phys. Rev. A
\textbf{11}, 473 (1975).



\bibitem {Heckel}B. Graner, Y. Chen, E. G. Lindahl, and B. R. Heckel, Phys.
Rev. Lett. \textbf{116}, 161601 (2016).




\bibitem {Axion}J.E.Kim, Phys. Rep. 150, 1 (1987); H.-Y. Cheng, Phys. Rep.
158, 1 (1988); J.E.Kim and G. Carosi, Rev. Mod. Phys. 82, 557 (2010); G.
Pignol, Int. J. Mod. Phys. A \textbf{30}, 1530048 (2015); Y.\thinspace V.
Stadnik and V.\thinspace V. Flambaum, Phys. Rev. D 89, 043522 (2014);
B.\thinspace M. Roberts, Y.\thinspace V. Stadnik, V.\thinspace A. Dzuba,
V.\thinspace V. Flambaum, N. Leefer, and D. Budker, Phys. Rev. D 90, 096005 (2014).



\bibitem {Barr}S. M. Barr, Phys. Rev. D \textbf{45}, 4148 (1992).


\bibitem {AtMolPhys}B. Bederson and H. Walther, \textit{Advances in Atomic,
Molecular and Optical Physics}, Vol. 40 (Academic Press, London, 1999).


\bibitem {Bouchiat}M. A. Bouchiat, and C. C. Bouchiat, Phys. Lett. \textbf{B48}, 111 (1974).




\bibitem {PNC2}I. B. Khriplovich and S. K. Lamoreaux, \textit{CP Violation
Without Strangeness}, (Springer, Berlin 1997).




\bibitem {AtPol}C. Bouchiat, Phys. Lett. B\textbf{57}, 284 (1975); E. A.
Hinds, C. E. Loving, and G. P. H. Sandars, Phys. Lett. B\textbf{62}, 97
(1976); A. M. Martensson-Pendrill, Phys. Rev. Lett. \textbf{54}, 1153 (1985);
V. A. Dzuba, V. V. Flambaum, and P. G. Silvestrov, Phys. Lett. B\textbf{154},
93 (1985); X.-G. He and B.
McKellar, Phys. Lett. B\textbf{390}, 318 (1997).



\bibitem {Neuffer}D. V. Neuffer and E. D. Commins, Phys. Rev. A \textbf{16}, 844 (1977); \textbf{16}, 1760 (1977); W. R. Johnson, Phys. Scr. \textbf{36}, 765 (1987).

\bibitem {YamanakaHg} N. Yamanaka, Phys. Rev. D \textbf{85}, 115012 (2012); K. Yanase, N. Yoshinaga, K. Higashiyama, and N. Yamanaka,
Phys. Rev. D\textbf{99}, 075021 (2019).

\bibitem {Colladay}V. Alan Kostelecky, Phys. Rev. D\textbf{69}, 105009 (2004); D. Colladay and V. A. Kostelecky, Phys. Rev. D \textbf{55},
6760 (1997); D. Colladay and V. A. Kostelecky \ Phys. Rev. D \textbf{58},
116002 (1998); S.R. Coleman and S.L. Glashow, Phys. Rev. D \textbf{59}, 116008 (1999).

\bibitem {fermion}V.A. Kostelecky and C. D. Lane, J. Math. Phys. \textbf{40},
6245 (1999); R. Lehnert, J. Math. Phys. \textbf{45}, 3399 (2004);\ D. Colladay
and V. A. Kostelecky, Phys. Lett. B \textbf{511}, 209 (2001); O. G. Kharlanov
and V. Ch. Zhukovsky, J. Math. Phys. \textbf{48}, 092302 (2007); R. Lehnert,
Phys. Rev. D \textbf{68}, 085003 (2003); V. A. Kostelecky and R. Lehnert,
Phys. Rev. D\textbf{\ 63, }065008 (2001)\textbf{;} S. Chen, B. Wang, and R.
Su, Classical Quantum Gravity \textbf{23}, 7581 (2006); B. Gon\c{c}alves, M.
M. Dias Junior, and B. J. Ribeiro, Phys. Rev. D \textbf{90}, 085026 (2014).

\bibitem {CPT}R. Bluhm, V.A. Kostelecky, and N. Russell, Phys. Rev. Lett.
\textbf{79}, 1432 (1997); Phys. Rev. D \textbf{57}, 3932 (1998); Phys. Rev.
Lett. \textbf{82}, 2254 (1999); R. Bluhm, V.A. Kostelecky, C. D. Lane, and N.
Russell, Phys. Rev. Lett. \textbf{88}, 090801 (2002); R. Bluhm and V.A.
Kostelecky, Phys. Rev. Lett.\textbf{\ 84}, 1381 (2000); R. Bluhm, V.A.
Kostelecky, and C. D. Lane, Phys. Rev. Lett. \textbf{84}, 1098 (2000); V.A.
Kostelecky and C.D. Lane, Phys. Rev. D \textbf{60}, 116010 (1999).

\bibitem {fermion2}T. Mariz, J. R. Nascimento, A. Yu. Petrov, Phys. Rev. D
\textbf{85}, 125003 (2012);  A. P.
Baeta Scarpelli, Marcos Sampaio, M. C. Nemes, B. Hiller, Eur. Phys. J. C
\textbf{56}, 571 (2008); F.A. Brito, L.S. Grigorio, M.S. Guimaraes, E. Passos,
C. Wotzasek, Phys.Rev. D \textbf{78}, 125023 (2008); F.A.Brito, E. Passos, and
P.V. Santos, Europhys. Lett. \textbf{95}, 51001 (2011); C. F. Farias, A. C.
Lehum, J. R. Nascimento, and A. Yu. Petrov, Phys. Rev. D \textbf{86}, 065035
(2012); J. R. Nascimento, A. Yu. Petrov, C. Wotzasek, and C. A. D. Zarro,
Phys. Rev. D \textbf{89}, 065030 (2014).

\bibitem {fermion3} R.V. Maluf, J.E.G. Silva, W.T. Cruz, and C.A.S. Almeida, Phys. Lett. B \textbf{738}, 341 (2014); B. Gonçalves, M.M. Dias Jr., B.J. Ribeiro, Phys. Rev. D \textbf{99} (2019) 096015;
	Z. Xiao, Phys. Rev. D \textbf{94}, 115020 (2016);
	Z. Xiao, Phys. Rev. D \textbf{93}, 125022 (2016);
	Z. Xiao, Bo-Qiang Ma, Int. J. Mod. Phys. A 24, 1359 (2009).


\bibitem {KM1}V. A. Kostelecky and M. Mewes, Phys. Rev. Lett. \textbf{87},
251304 (2001); V. A. Kostelecky and M. Mewes, Phys. Rev. D\textbf{\ 66},
056005 (2002); V. A. Kostelecky and M. Mewes, Phys. Rev. Lett. \textbf{97},
140401 (2006); C A. Escobar, M. A. G. Garcia, Phys.Rev. D\textbf{92}, 025034 (2015); A.~Mart\'{\i}n-Ruiz and C.A.~Escobar, Phys. Rev. D~\textbf{94},
076010 (2016).


\bibitem {CFJ}S. M. Carroll, G. B. Field, and R. Jackiw, Phys. Rev. D 41,
1231 (1990); C. Adam and F. R. Klinkhamer, Nucl. Phys.
B607, 247 (2001); B657, 214 (2003); Y.M. P. Gomes and
P. C. Malta, Phys. Rev. D 94, 025031 (2016).

\bibitem {photons1} A. P. Baeta \ Scarpelli, H.
Belich, J. L. Boldo, J.A. Helayel-Neto, Phys. Rev. D \textbf{67}, 085021
(2003); L.H.C. Borges, F.A. Barone, and J.A. Helayel-Neto, Eur. Phys. J. C
\textbf{74}, 2937 (2014); T. R. S. Santos and R. F. Sobreiro, Phys. Rev. D
\textbf{91}, 025008 (2015); T. R. S. Santos and R. F. Sobreiro, Braz. J. Phys. 46, 437 (2016)

\bibitem {YM}T. R. S. Santos, R. F. Sobreiro, A. A. Tomaz, Phys. Rev. D 94, 085027 (2016); Tiago R. S. Santos, Rodrigo F. Sobreiro, Eur. Phys. J. C 77, 903 (2017).

	
\bibitem {Casimir} O. G. Kharlanov, V. Ch. Zhukovsky, Phys.Rev.D81, 025015 (2010); A. Mart\'{\i}n-Ruiz, C.A. Escobar, Phys.Rev. D \textbf{94}, 076010 (2016); B. Cruz, E. R. Bezerra de Mello, A. Yu. Petrov, Phys.Rev. D96, 045019 (2017); C.A. Escobar, J.Phys.Conf.Ser. 952, 012010 (2018);  A. Mart\'{\i}n-Ruiz, C.A. Escobar, Phys. Rev. D 95, 036011 (2017)

\bibitem {Vertex}F.R. Klinkhamer and M. Schreck, Nucl. Phys. \textbf{B848}, 90
(2011); M. Schreck, Phys. Rev. D \textbf{\ 86}, 065038 (2012); M. A. Hohensee,
R. Lehnert, D. F. Phillips, and R. L. Walsworth, Phys. Rev. D \textbf{80},
036010 (2009); A. Moyotl, H. Novales-S\'{a}nchez, J. J. Toscano, E. S. Tututi,
Int. J. Mod. Phys. A \textbf{29}, 1450039 (2014); Int. J. Mod.Phys. A
\textbf{29}, 1450107 (2014); M. Cambiaso, R. Lehnert, R. Potting, Phys. Rev. D
\textbf{90}, 065003 (2014); R. Bufalo, Int. J. Mod. Phys. A \textbf{29},
1450112 (2014); G. P. de Brito,  P. C. Malta, and L. P. R. Ospedal, Phys. Rev. D \textbf{95}, 016006 (2017).

\bibitem {QED} G. P. de Brito, J. T. Guaitolini Jr, D. Kroff, P. C. Malta, and C. Marques, Phys. Rev. D \textbf{94} 056005 (2016); T. R. S. Santos, R. F. Sobreiro, Phys. Rev. D \textbf{94}, 125020 (2016); A. F. Santos, and F. C. Khanna, Adv. High E. Phys. 2018, 4596129 (2018); F.E.P. dos Santos, M. M. Ferreira Jr, Symmetry \textbf{10}, 302 (2018); T. P. Netto, Phys.Rev. D\textbf{97}, 055048 (2018); J. R. Nascimento, A. Yu. Petrov, Carlos M. Reyes, Eur. Phys. J. C \textbf{78}, 541 (2018). 

\bibitem {EW1}J. P. Noordmans, H. W. Wilschut, and R. G. E. Timmermans, Phys.
Rev. C \textbf{87}, 055502 (2013); Phys. Rev. Lett. \textbf{111}, 171601 (2013); B. Altschul,
Phys. Rev. D 87, 096004 (2013); Phys. Rev. D \textbf{88}, 076015 (2013); J. P.
Noordmans and K. K. Vos, Phys. Rev. D \textbf{89}, 101702(R) (2014); J. S. Diaz, V. A.
Kostelecky, and R. Lehnert, Phys. Rev. D \textbf{88}, 071902(R) (2013); J. S. Diaz,
Adv. High Energy Phys. 2014, 305298 (2014); K. K. Vos, H. W. Wilschut, and R.
G. E. Timmermans, Phys. Rev. C \textbf{91}, 038501 (2015); Phys. Rev. C \textbf{92}, 052501(R)
(2015); Rev. Mod. Phys. \textbf{87}, 1483 (2015).

\bibitem {EW2}J. Castro-Medina, H. Novales-Sanchez, J. J. Toscano, Int. J.
Mod. Phys. \textbf{A 30}, 1550216 (2015); M. A. L\'{o}pez-Osorio, E.
Mart\'{\i}nez-Pascual, J. J. Toscano, J. Phys. G: Nucl. Part. Phys. \textbf{43} (2016)
025003; J. I. Aranda, F. Ramirez-Zavaleta, D. A. Rosete, F. J. Tlachino, J. J.
Toscano, E. S. Tututi, Int. J. Mod. Phys. \textbf{A 29}, 1450180 (2014).

\bibitem {EW3} D. Colladay, J. P. Noordmans, R. Potting, Symmetry \textbf{9}, 248 (2017);  Phys.Rev. D\textbf{96}, 035034 (2017); J.Phys.Conf.Ser. \textbf{952}, 012021 (2018); C.A. Escobar, J.P. Noordmans, R. Potting, Phys.Rev. D\textbf{97}, 115030 (2018); A.I. Hern\'{a}ndez-Ju\'{a}rez, J. Monta\~{n}o, H. Novales-S\'{a}nchez, M. Salinas, J.J. Toscano, O. V\'{a}zquez-Hern\'{a}ndez, Phys.Rev. D\textbf{99}, 013002 (2019).


\bibitem {NMSME1}V.A.~Kosteleck\'{y} and M.~Mewes, Phys. Rev.
D~\textbf{80}, 015020 (2009); M.~Mewes, Phys. Rev. D~\textbf{85}, 116012
(2012); M.~Schreck, Phys. Rev. D~\textbf{89}, 105019 (2014).

\bibitem {NMSME2}V.A.~Kosteleck\'{y} and M.~Mewes, Phys. Rev.
D~\textbf{88}, 096006 (2013); M.~Schreck, Phys. Rev. D~\textbf{90}, 085025 (2014); J. A. A. S. Reis, M. Schreck, hys. Rev. D ~\textbf{95}, 075016 (2017).

\bibitem {Reyes}R. C. Myers and M. Pospelov, Phys. Rev. Lett. \textbf{90},
211601 (2003); C. M. Reyes, L. F. Urrutia, J. D. Vergara, Phys. Rev. D
\textbf{78}, 125011 (2008); Phys. Lett. B \textbf{675}, 336 (2009); C. M.
Reyes, Phys. Rev. D \textbf{82}, 125036 (2010); Phys. Rev. D \textbf{80},
105008 (2009); Phys. Rev. D \textbf{87}, 125028 (2013). C. M. Reyes, S.
Ossandon, and C. Reyes, Phys. Lett. B \textbf{746}, 190 (2015);  A. Celeste, T. Mariz, J. R. Nascimento, A. Yu. Petrov, Phys. Rev. D \textbf{93}, 065012 (2016).

\bibitem {HD}M. Cambiaso, R. Lehnert, and R. Potting, Phys. Rev. D
\textbf{85}, 085023 (2012); B. Agostini, F. A. Barone, F. E. Barone, P. Gaete,
and J. A. Helay\"{e}l-Neto, Phys. Lett. B \textbf{708}, 212 (2012); L.
Campanelli, Phys. Rev. D \textbf{90}, 105014 (2014); R. Bufalo, B.M. Pimentel,
and D.E. Soto, Phys. Rev. D \textbf{90}, 085012 (2014).

\bibitem {Ding}Y. Ding and V.A. Kostelecky, Lorentz-violating spinor
electrodynamics and Penning traps, Phys. Rev. D \textbf{94}, 056008 (2016).

\bibitem {NModd1}H.~Belich, T.~Costa-Soares, M.M.~Ferreira, Jr., and
J.A.~Helay\"{e}l-Neto, Eur. Phys. J. C~\textbf{41}, 421 (2005); H.~Belich,
L.P.~Colatto, T.~Costa-Soares, J.A.~Helay\"{e}l-Neto, and M.T.D.~Orlando, Eur.
Phys. J. C~\textbf{62}, 425 (2009); B.~Charneski, M.~Gomes, R.V.~Maluf, and
A.J.~da~Silva, Phys. Rev. D~\textbf{86}, 045003 (2012); A.F.~Santos, and
Faqir~C.~Khanna, Phys. Rev. D~\textbf{95}, 125012 (2017).

\bibitem {NModd2}G. Gazzola, H.G. Fargnoli, A.P. Baeta Scarpelli, M. Sampaio, M.C. Nemes, J. Phys. G 39, 035002 (2012); A.P. Baeta Scarpelli, J. Phys. G 39, 125001 (2012);
L.C.T. Brito, H.G. Fargnoli, A.P. Baeta Scarpelli, Phys. Rev. D 87, 125023 (2013); K. Bakke, H. Belich, E.O. Silva, J. Math. Phys. 52, 063505 (2011);
J. Phys. G 39, 055004 (2012); Ann. Phys. (Leipz.) 523, 910 (2011);  Y.M.P. Gomes, J.T. Guaitolini Jr, Phys. Rev. D~\textbf{99}, 055006 (2019).

\bibitem {PetrovPLB2016} L. H. C. Borges, A. G. Dias, A. F. Ferrari, J. R. Nascimento, and A. Yu. Petrov, Phys. Rev. D \textbf{89}, 045005 (2014); L.H.C.~Borges, A.G.~Dias, A.F.~Ferrari,
J.R.~Nascimento, and A.Yu.~Petrov, Phys. Lett. B~\textbf{756}, 332 (2016); A.J.G.Carvalho, A.F.Ferrari, A.M.de Lima, J.R.Nascimento, and A.Yu.Petrov, Nucl. Phys. B \textbf{942}, 393 (2019).

\bibitem {Victor}V. E. Mouchrek-Santos and M. M. Ferreira, Jr.,
Phys. Rev. D \textbf{95}, 071701(R) (2017); J.Phys.Conf.Ser. 952, 012019 (2018). 
\bibitem {Anacleto} M.A. Anacleto, F.A. Brito, E. Maciel, A. Mohammadi, E. Passos, W.O. Santos, and J.R.L. Santos, Phys. Lett. B~\textbf{785}, 191 (2018).
\bibitem {Anacleto2} E. Passos, M.A. Anacleto, F.A. Brito, O. Holanda, G.B. Souza, and C.A.D. Zarro, Phys. Lett. B~\textbf{772}, 870 (2017).

\bibitem {Haghig}M. Haghighat, I. Motie, and Z. Rezaei, Int. J. Mod. Phys. A
\textbf{28}, 1350115 (2013).

\bibitem {Pospelov2008}P. A. Bolokhov, M. Pospelov, and M. Romalis, Phys. Rev.
D \textbf{78}, 057702 (2008).

\bibitem {FredeNM1}R. Casana, M.M. Ferreira, Jr., E. Passos, F. E. P. dos
Santos, and E. O. Silva, Phys. Rev. D \textbf{87}, 047701 (2013).

\bibitem {Jonas1}J. B. Araujo, R. Casana and M.M. Ferreira, Jr., Phys. Rev. D
\textbf{92}, 025049 (2015).

\bibitem {Jonas2}J. B. Araujo, R. Casana and M.M. Ferreira, Jr., Phys.Lett.
B760, 302-308 (2016).

\bibitem {Jonas3}J. B. Araujo, R. Casana and M.M. Ferreira, Jr., Phys. Rev. D
\textbf{97}, 055032 (2018).



\bibitem {Stadnik}Y.V. Stadnik, B.M. Roberts, and V.V. Flambaum, Tests of
\ CPT and Lorentz symmetry from muon anomalous magnetic dipole moment, Phys.
Rev. D \textbf{90}, 045035 (2014).

\bibitem {Gomes}A. H. Gomes, V.A. Kostelecky, and A. J. Vargas, Phys. Rev. D\textbf{90}, 076009 (2014).

\bibitem {Haghig2}S. Aghababaei, M. Haghighat, I. Motie, Muon anomalous
magnetic moment in the standard model extension, Phys.Rev. D96, 115028 (2017).


\bibitem {LVArbDim}V. Alan Kostelecky, and Z. Li, Phys. Rev. D\textbf{99}, 056016 (2019).



\bibitem {Sternheimer1}R. M. Sternheimer, Phys. Rev. \textbf{127}, 4, (1962).



\bibitem {PNC}I. B. Khriplovich, \textit{Parity Nonconservation in Atomic
Phenomena}, 1st Edition (CRC Press, Philadelphia, 1991).




\bibitem {Sidereal}R. Bluhm, V. Alan Kostelecky, C. D. Lane, and N. Russel,
Phys. Rev. Lett. \textbf{88},\ 090801 (2002); Phys. Rev. D \textbf{68},
125008\ (2003); V.A. Kostelecky and\ M.\ Mewes,\ Phys.\ Rev.\ D\ \textbf{66},\ 056005\ (2002).







\end{thebibliography}
\end{document}